\documentclass{article}
\usepackage{spconf,amsmath,graphicx}

\usepackage{booktabs}

\title{Cross-Modal Fusion Techniques for Utterance-Level Emotion Recognition from Text and Speech}
\name{Jiachen Luo$^1$, Huy Phan$^{2*}$\thanks{$^{*}$The work was done when H. Phan was at Centre for Digital Music, Queen Mary University of London, UK and prior to joining Amazon.}, Joshua Reiss$^1$\thanks{Thanks to the China Scholarship Council and Queen Mary University of London for funding.}}
\address{Centre for Digital Music, Queen Mary University of London, UK\\$^2$Amazon Alexa, Cambridge, MA, USA}
\begin{document}
\ninept
\maketitle
\begin{abstract}
Multimodal emotion recognition (MER) is a fundamental complex research problem due to the uncertainty of human emotional expression and the heterogeneity gap between different modalities. Audio and text modalities are particularly important for a human participant in understanding emotions. Although many successful attempts have been designed multimodal representations for MER, there still exist multiple challenges to be addressed: 1) bridging the heterogeneity gap between multimodal features and model inter- and intra-modal interactions of multiple modalities; 2) effectively and efficiently modeling the contextual dynamics in the conversation sequence. In this paper, we propose Cross-Modal RoBERTa (CM-RoBERTa) model for emotion detection from spoken audio and corresponding transcripts. As the core unit of the CM-RoBERTa, parallel self- and cross- attention is designed to dynamically capture inter- and intra-modal interactions of audio and text. Specially, the mid-level fusion and residual module are employed to model long-term contextual dependencies and learn modality-specific patterns. We evaluate the approach on the MELD dataset and the experimental results show the proposed approach achieves the state-of-art performance on the dataset.
\end{abstract}


\section{Introduction}
\label{sec:intro}

Emotion recognition in conversations (ERC) is vital and very challenging in the natural human machine interaction [1], intelligent education tutoring [2], and mental health analysis applications [3]. In daily life, humans utter a multi-turn conversation in a natural way which conveys emotion state through language and nonverbal content (e.g., facial expression and body language) [4]. Different modalities all carry emotion-relevant information and how to efficiently and effectively fuse heterogeneous inputs across multiple modalities has been an active research focus. 

In daily life, text modality is often accompanied by audio modality. The sentiment content contained in text features is complemented by the multiple information such as pitch, intensity, pause, loudness, and other frequency-related measures [5,6]. The effective interaction between text and audio data enables more comprehensive content and provides more emotional information [7$\sim$9]. As the example shown in Fig. 1, a single word ``Okay'' is ambiguous, and it can express different emotions. It is challenging to determine the associated emotion according to one turn of textual utterance. After introducing corresponding low voices and sobs, it is not difficult to discern the sentiment of this sentence is negative. However, fusing different modalities in an effective manner is a non-trivial task that researchers often have to face.

\begin{figure}[htb]

\begin{minipage}[b]{1.0\linewidth}
  \centering
  \centerline{\includegraphics[width=8.5cm]{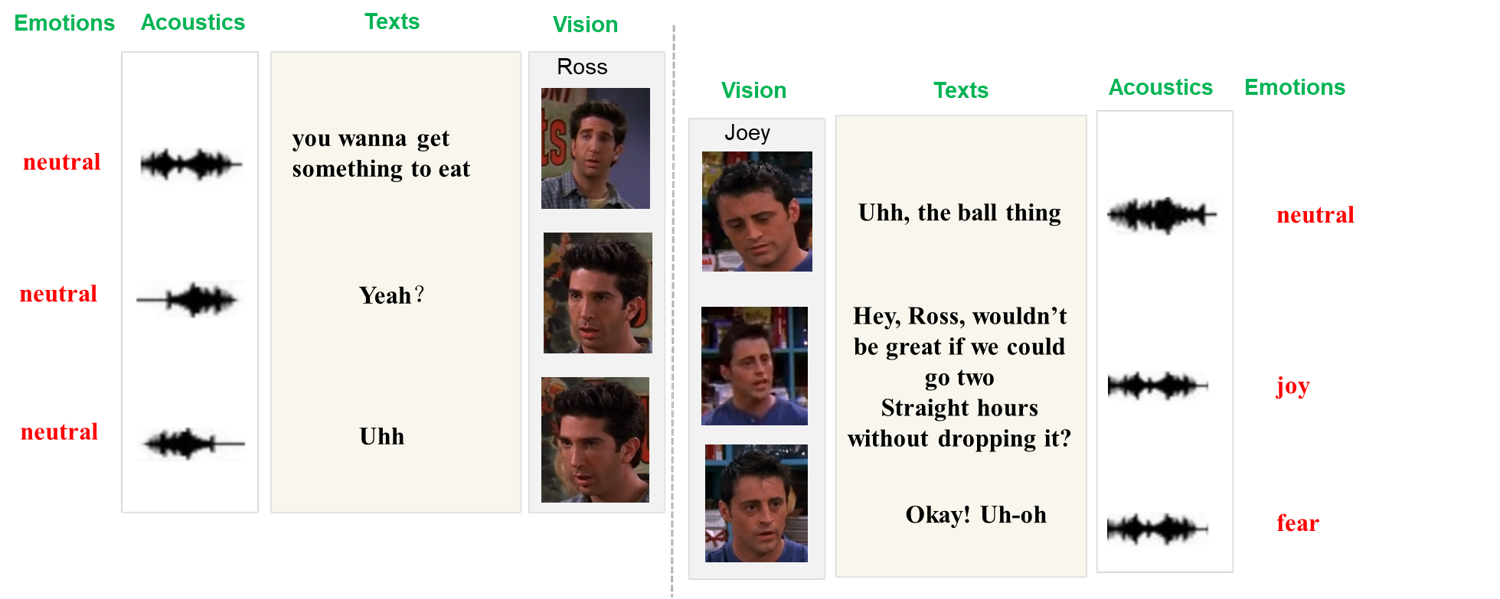}}
\end{minipage}
\caption{\small Emotion dynamic of speakers in a dialogue in comparison.}
\label{fig:res2}
\end{figure}

Previous methods have achieved good performance [8,9], there still exist key challenges in multimodal emotion recognition:  different modalities depend on independent preprocessing and feature excavation designs due to heterogeneous space; and to make an applicable and generalizable model for both the individual modalities and fusion model, it is necessary to learn intra- and cross-modal interactions to reveal discriminative emotion content; and  emotion is a subjective concept [10$\sim$12]. Moreover, textual information and the associated context conveys more influential information in inferring the speaker’s emotions [13], and plays a crucial role in inferring emotions in conversations. Unfortunately,
the existing approaches cannot efficiently capture the textual emotional-relevant content in the fusion processing, and often the learning of intra- and cross-modal information will
lose some semantic content [12,14].

To tackle the above problems, we proposed a Cross-Modal RoBERTa to integrate the information of audio modality in fine-tuning the pre-trained RoBERTa extractor in our architecture. As the unit of the CM-RoBERTa, cross- and self-attention layers were designed to learn inter- and intra-modality information, and could adaptively extend the interactions and correlations of information from different sources. In addition, we designed a temporal calibration module to effectively preserve the local information (text and audio) from a single modality and the overall information generated from the contextual level. We evaluate it on the public multimodal sentiment dataset MELD [15]. The experiment results showed a significant improvement in model performance and efficiency. The main contributions of this paper could be summarized as follows:

$\bullet$ We propose a CM-RoBERTa model that introduces the information audio modality to help text-modality fine-tuning of the pre-trained RoBERTa model.

$\bullet$ We employ parallel self- and cross-attention that can dynamically adjust the attentive weights through the interaction between audio and text and capture inter- and intra-modal information.

$\bullet$ We utilize mid-level fusion and residual connection to capture long-term contextual dependencies, and learn modality-specific patterns.

$\bullet$ We show that our proposed approach is superior to state-of-the-art methods for conversational emotion recognition.

\section{Related Works}
\label{sec:format}

Emotion recognition in conversations is a popular research area because of their applications in various areas. Since human emotions are expressed in many ways, and there are some limitations in the recognition of emotions by single modal, multimodal emotion recognition has gradually become a new trend. A basic challenge for multimodal emotion recognition is how to integrate the multimodal information effectively. 

To date, multimodal emotion recognition is roughly divided into four categories [14$\sim$21]: early fusion, late fusion, model-level fusion, and hybrid fusion. Early fusion is to fuse input features from different modalities through concatenating and other ways. Poria et al. leveraged  context-aware utterance representations for emotion recognition [15]. Due to the complex inter-modality dynamics at the input level, this feature fusion approach was incapable of capturing the intra-modality information. Late fusion, instead, trains an independent classification model of each modality and utilizes decision voting or other classifier combination strategies. Xie et al. employed crossmodal transformer fusion for audio, text and video late-level fusion, avoiding the difficulty of fusing heterogeneous information [22]. However, such approaches ignored associations across multiple modalities and failed to model mutual correlations. 

On the other hand, model-level fusion is a compromise between the former two where the fusion happens between the intermediate representations of the multimodal features. Akhtar et al. designed a context-level inter-modal attention module for learning the joint association between textual-audio-visual information of utterances [20]. Hybrid fusion combines early fusion and late fusion. Wollmer et al. fused audio-video information with Bi-LSTM for audio-visual emotion recognition[14]. How to design an effective and efficient structure to capture inter- and intra- modalities features is necessary in multimodal emotion recognition. 

With the popularity of the attention mechanism, it has recently attracted some in the field for its ability to fuse multimodal information. Devamanyu et al. used a multi-attention recurrent network framework that learned multimodal interactions [23]. Also, recently transfer learning techniques that utilized pre-trained networks for feature extraction have become popular [24]. RoBERTa, a BERT-based model, has achieved better performance by fine-tuning from pre-trained weights for emotion recognition tasks [25]. In our work, we proposed an effective and efficient fusion approach for fine-tuning by fusing heterogeneous nonverbal information (audio) that complements the linguistic expressions of RoBERTa.

\section{Methodology}
\label{sec:typestyle}
We leveraged the multi-modal and contextual information for discerning the discriminative emotion for an utterance. The architecture of the model is shown in Fig. 2.

\begin{figure*}[htb]

\begin{minipage}[b]{1.0\linewidth}
  \centering
  \centerline{\includegraphics[width=18cm]{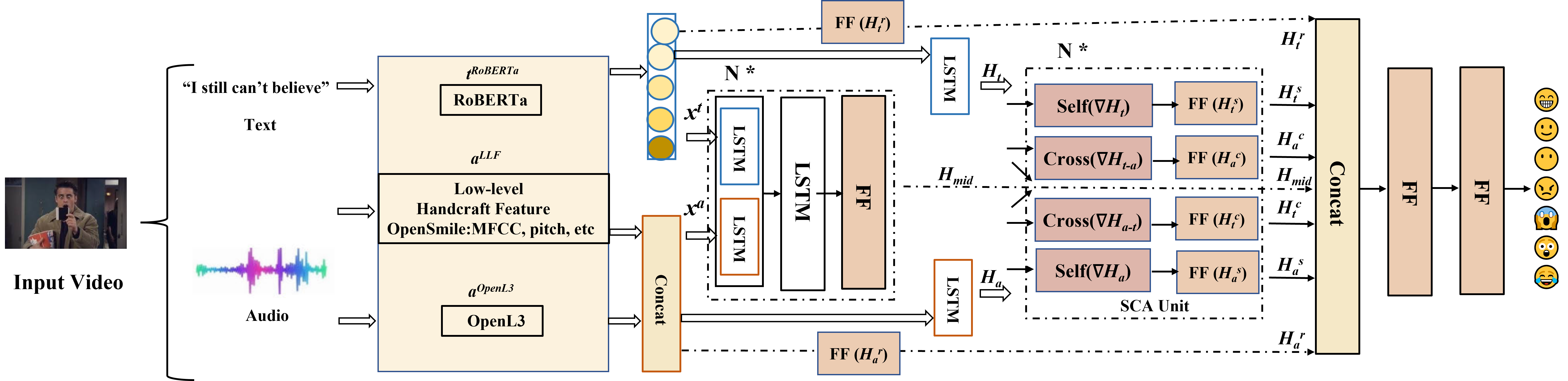}}
\end{minipage}
\caption{\small Overview architecture of the Cross-Modal RoBERTa Fusion Network N represents two layers, and the first two parallel LSTM are exactly the same as the last two parallel LSTM.}
\label{fig:res1}
\end{figure*}

\subsection{Audio and Text Feature Extraction}
\label{ssec:subhead3}
The input of the CM-RoBERTa model consists of the effective heterogeneous representation (handcrafted and bottleneck features) for each audio and text from the data (see Fig. 2). For each audio utterance, we extract 6552 low-level handcrafted features ($a^{LLF}$) constituting several descriptors and various statistical functionals of varied vocal and prosodic features using the openSMILE toolkit [26]. In addition, a pre-trained OpenL3 model [27] is employed to extract bottleneck embedding ($a^{OpenL3}$). $a^{LLF}$ and $a^{OpenL3}$ are then concatenated to obtain audio presentation $x^a = (a^{LLF} {\oplus} a^{OpenL3})$. Where ${\oplus}$ denotes concatenation.
For text modality, the RoBERTa encoder ($t^{RoBERTa}$) received the transcript of the utterance to produce activations from the final four layers. Then, these four were concatenated to obtain the context-independent utterance feature vector $x^t$ with a dimension of 4096 [25].  

\subsection{Cross-Modal RoBERTa}
\label{ssec:subhead4}
As shown in Fig. 2, CM-RoBERTa first uses a bidirectional LSTM to encode acoustic and textual features, denoted as $H_a$ and $H_t$, respectively. We input them ($H_a$ and $H_t$) into the self- and cross-attention unit (SCA) to capture intra- and inter- modal interactions between each pair of acoustic frames and textual words. The SCA module consists of two stacked parallel cross- and self-attention layers and feed-forward layers. The main insight of the module is to learn modality associations and modality-specific patterns, and then propagate information from intra- and inter-modal perspectives according to the attentive weights. 

Following [28], we estimate the associations between audio and text modalities in a crossed way, whose query ($Q_m$), key ($K_m$), value ($V_m$) are the representation of modality $m$, $H_m$, under different projection space, where $m$ is either audio ($a$) or text ($t$). The specific formulae are as follows: 
\begin{align} 
\nabla H_{a-t} &= softmax(Q_aK^\mathsf{T}_t/\sqrt{d})V_t, \\
\nabla H_{t-a} &=  softmax(Q_tK^\mathsf{T}_a/\sqrt{d})V_a,
\end{align}
where $\nabla H_{a-t}$ and $\nabla H_{t-a}$ represent the propagated information from audio to text and text to audio, respectively. $d$ denotes the key dimensionality. 
Later, we update the features of one modality with the information propagating from the other modality:
\begin{align} 
H^{L^N_a} &= LayerNorm(H_a {\oplus} \nabla H_{t-a}), \\
H^{L^N_t} &=  LayerNorm(H_t {\oplus} \nabla H_{a-t}).
\end
{align} 
To further enhance the representation capacity, we feed the learned weighed information into the full connection to obtain the final cross-attention information between audio and text modalities:
\begin{align} 
H^{c_a} &= LayerNorm(H^{L^N_a} {\oplus} \nabla  FeedForward(H^{L^N_a})), \\
H^{c_t} &= LayerNorm(H^{L^N_t} {\oplus} \nabla  FeedForward(H^{L^N_t})).
\end{align} 
We denote the final cross-attention information between audio and text modalities as $H^c_a$ and $H^c_t$, respectively. Where $\oplus$ is concatenation operation. ”

The self-attention mechanism has the similar principle with the cross-attention mechanism. The only different is the query, key, and value are from the same modality. The whole process for the one stacked layer in the self-attention module can be summarized as: 
\begin{align} 
\nabla H_m &= softmax(Q_mK^\mathsf{T}_m/\sqrt{d})V_m, \\
H^{LN_s}_m &= LayerNorm(H_m {\oplus} \nabla  H_m), \\
H^s_m &= LayerNorm(H^{LN_s}_m {\oplus} \nabla  FeedForward(H^{LN_s}_m)),
\end{align} 
where $\nabla H_m$ is the propagated information within modality $m$, $m \in \{a, t\}$. We denote the outputs of the last stacked layer in two self-attention modules as $H^s_a$ and $H^s_t$, respectively. In parallel, we then perform mid-level fusion of the joint features with modality-specific patterns. They help in capturing the inter- and intra-modality dynamics. Individual modalities $(x_a, x_t)$ are passed in parallel through two stacked LSTM layers separately to store single modality feature $(H^m_a, H^m_t)$ efficiently, then combined via a bidirectional LSTM to explore modality associations:
\begin{align} 
H_{mid} = FeedForward(LSTM([H_a, H_t])).
\end{align} 

Following prior works [13], we utilized a residual connection on audio and textual representation to keep the original structure of the data $(H^r_a, H^r_t)$. Then it is passed through a linear and a normalization layer. Finally, we concatenate them $(H^c_a, H^c_t, H^s_a, H^s_t, H_{mid}, H^r_a, H^r_t)$ as the aggregating representation and presented it to two fully-connected layers for classification. 


\section{Experiments}
\label{sec:majhead}

\subsection{Dataset and Feature Sets}
\label{ssec:subhead6}
\subsubsection{Dataset}
\label{sssec:subsubhead}

Multi-modal Emotion Dataset (MELD) is a multi-modal (visual, audio, and text) and multi-speaker emotion recognition dataset consisting of dialogues from the Friends TV show [15]. MELD has more than 13000 utterances and 1400 conversations. Each utterance is annotated using the following discrete categories: anger, joy, sadness, neutral, disgust, fear and surprise. The dataset is split into the training set, validation set and test set which contains 9989, 1109, and 2610 utterances [15]. Weighted F1-score was used to evaluate the performance, as the class distribution is highly imbalanced.

\subsubsection{Audio Features}
The audio features consisted of two types of features for each utterance: handcrafted and bottleneck features. As handcraft features, we extracted 6552 features using the openSMILE toolkit [26]. They encapsulated features like MFCC, pitch, etc. In addition, we extracted deep-learning-based bottleneck features using the OpenL3 model [27]. 
OpenL3 are self-supervised embeddings learned from AudioSet [27]. The OpenL3 network contains four blocks stacked together, each block consists of three convolutional layers and one pooling layer.
We processed audio data to conform to the network input and extracted 6144 high-level bottleneck features. 

\subsubsection{Textual Features}
We fine-tuned the pre-trained RoBERTa Large model (Uncased: hidden-1024, heads-24, layer-24) as our language model for extracting the features from the text modality [25]. RoBERTa took the transcript of the utterance as input and produced rich contextual textual representation from the final four layers. This produced four 1024-dimension vectors for every token in the input. Then, we concatenated them to obtain the context independent utterance feature vector with a dimension of 4096.

\begin{table*}[htbp]
	\centering
	{ Table 1: Overall performance comparison between the proposed method, the state-of-the-art, and the baselines. The weighted F1-scores were used as the metric. Note that our unimodal audio-based and text-based models are comprised of LSTM layer, and residual block.}
	\begin{tabular}{l c c c c c}
		\hline
		model & audio & text & bimodal & bimodal - text & reference \\
		\hline
		SMIN & 45.2 & 63.65 & 64.5 & 0.85 & [29] \\
		EmoCaps & 31.26 & 63.51 & 63.73 & 0.22 & [17] \\
		EmbraceNet & 40 & 61 & 63.1 & 2.1 & [22] \\
		CMN & 38.3 & 54.5 & 55.5 & 1 & [23] \\
		ICON & 37.7 &  54.6 & 56.3 & 1.7 & [30] \\
		DialogueRNN & 34 & 55.1 & 57 & 1.9 & [16] \\
		bc-LSTM & 36.4 & 54.3 & 56.8 & 2.5 & [15] \\
	\textbf{Proposed method without SCA unit} & \textbf{43.10} & \textbf{63.83} & \textbf{65.66} & \textbf{1.83}\\
		\textbf{Proposed method} & \textbf{43.10} & \textbf{63.83} & \textbf{66.28} & \textbf{2.45} & \\
	
		\hline
	\end{tabular}
	\label{tab:table1}
\end{table*}

\begin{table*}[htbp]
	\centering
	{ Table 2: The class-wise weighted F1 scores on the MELD dataset}
	\begin{tabular}{l c c c c c c c c}
		\hline
		Proposed method & neutral & surprise & fear & sadness & joy & disgust & anger & w-average F1 \\
		\hline
		audio modality & 62.9 & 20.12 & 0.0 & 6.7 & 22.65 &0.0 & 53.21 & 43.10 \\
		text modality & 76.39 & 60.47 & 30.5 & 42.21 & 63.21 & 31.74 & 45.82 & 63.83 \\
		bimodality & 78.67 & 63.42 & 28.7 & 42.19 & 63.65 & 30.91 & 53.56 & 66.28 \\
		\hline
	\end{tabular}
	\label{tab:table2}
\end{table*}

\subsection{Baselines}
\label{ssec:subhead1}
For a comprehensive evaluation, we compared our model with the following baselines and state-of-the-art models in multi-modal emotion recognition (Table 1). \\
$\bullet$ SMIN integrated semi-supervised learning with multimodal interactions for conversational emotion recognition [29]. $\bullet$ EmoCaps utilized the transformer structure for contextual emotion classification [17]. $\bullet$ EmbraceNet employed transformer-based crossmodality fusion to estimate the emotion [22]. $\bullet$ CMN used attention-based hops to capture inter-speaker dependencies [23]. $\bullet$ ICON employed a memory network to store contextual summaries for classification [30]. $\bullet$ DialogueRNN employed three GRUs to model the speaker, the context and the emotion of the preceding utterances [16]. $\bullet$ bc-LSTM performed context-dependent sentiment analysis and emotion recognition [15].

\subsection{Experimental Setup}
\label{ssec:subhead2}
The model is trained with batch-size 32 for 200 epochs and stop training if the validation loss does not decrease for 15 consecutive epochs. The number of hidden neurons in the model is 128. We employed Adam optimizer with the learning rate of 1e-4. The \emph{L}$_2$ weight decay was set to 3e-4. Cross-entropy loss  with softmax activation in the last layer was used for network training.

\section{Results and Discussion}
\label{sec:print}

Table 1 summarizes the performance of unimodal and multimodal variants of the state-of-the-art text-based ERC systems along with the proposed CM-RoBERTa model. It should be highlighted that the text modality performs noticeably better than the audio modality (see Tables 1 and 2). However, for the MELD dataset, several factors make emotion detection in conversations using audio considerably hard. First, there is a lot of background noise. Second, the length of utterances is shorter than other benchmark datasets (e.g., IEMOCAP [31]). What’s worse, there are more than 5 speakers in most of the conversations.

Text tends to effectively alleviate the above problems by providing rich-emotional features in the joint representation. The word-level time-dependent textual content carries a wealth of information, such as topic, viewpoint, intent, and so on. This knowledge is essential to capture the nature and flow of the emotional dynamics of speakers. For short communication, a specific word (e.g., amazing) may directly represent the emotional state and dominate the final decision. In our work, we employed the fine-tuned RoBERTa model to extract textual utterance level features. Compared with other text embedding (see Table 2), RoBERTa attains marginally better results [25], which may be attributed to dynamically changing mask pattern and training the model longer with more data, enabling better marriage with the large-scale language model.

However, emotion recognition in conversations sometimes is dependent to its audio, rather than text, to infer what the emotion of a participant in conversations. We notice the high-level deep learning based acoustic features-openL3 embeddings play significantly better than low-level traditional features: pitch, MFCC, and other audio features in discriminating emotions in conversations (Table 3). For the 7-class task, the performance of our model with pre-trained openL3 audio embedding is 4.9\% than the baseline bc-LSTM model. The pre-trained OpenL3 model is useful in limited and unbalanced MELD dataset as knowledge transfer should result in better high-level emotion audio features. In real-word scenarios with the presence of noise and multiple speakers, pitch, MFCCs might not be ideal for emotion recognition in conversations. Additionally, in our work, by combing low-level descriptors and high-level OpenL3 acoustic representation, surprise and anger can gain 2.95\%, and 7.74\% on F1-score, respectively, over the text-based unimodal model (Table 2). It is showed that anger is best recognized from speech and is characterized by high pitch and energy. 

Introduction of the audio modality’s information to text modality can empower significant advantages, as it allows for improved emotion detection by helping to fuse additional audio features, disambiguate linguistic meaning and bridge a gap to real-world environments. Multimodality extracts and fuses vital information from the respective modality source and captures richer representation and performance than the individual modalities. The inter, intra and cross-modal interactions, correlations and relationships between multiple modalities are release more emotion-relevant content for better performance and prediction. 

\begin{center} 
{\small Table 3:Ablation audio features study on the MELD dataset}
\end{center}
\begin{center}
 \begin{tabular}{l c c c c}
 \hline
 
 Method &  w-average F1 \\ 
 \hline
 \emph{w/o} low-level descriptors  & 37.3 \\
 
\emph{w/o} openL3 embedding & 41.3  \\

 our audio features & 43.1 \\

 \hline
\end{tabular}
\end{center}

Our proposed multimodal fusion framework gained 2.45\% on F1-score over the text-based unimodal model, and SCA unit improvements about 1.83\% on F1-score (see Table 1). The superior performances of CM-RoBERTa might benefit for the following reasons. First, the early multi-modal cross-attention fused the underlying temporal interaction between text and audio features. We introduced cross-attention calibration by temporal-awareness to deduce the mutually exclusive and dependent between heterogeneous knowledge. By giving more weight to relevant audio modality, we could introduce the information of the audio modality to help the text modality effectively adjust the weight of words. We speculate that our cross-attention fusion unit can effectively employ the different knowledge into a new space to strengthen the details embedded in a single modality and adaptively fuse the implicit complementary content to magnify the interactions and correlations. Second, the self-attention mechanism was also employed to propagate information within each modality. By modeling the intra-modal interactions, we can capture learn modality-specific patterns, and distribute the self-attention weights over the entire dialog sequence for each modality. Third, mid-level fusion and residual blocks were targeted at explicitly modeling both inter- and intra-modal interaction of audio and text. This aspect has the power to fuse heterogeneous space, retain contextual intra-modality-specific representation and facilitate synchronization.

\section{Conclusion}
\label{sec:page}

In this paper, we proposed the CM-RoBERTa model that effectively fused multi-modal emotion recognition using a fine-tuned RoBERTa model for languages and heterogeneous features unification for handcrafted and bottleneck audio features. Thanks to the parallel self- and cross-attention modules, mid-level fusion and residual blocks that could explicitly model inter- and intra-modal interactions within/between audio and text, capture long-term contextual dependencies, and learn modality-specific patterns. They achieved state-of-the-art utterance-level recognition performance when evaluated on the MELD dataset.

There are some limitations to our method. The work relied on the controlled database with non-realistic recording conditions and error-less text transcriptions. Thus, in practice, more robust models need to be designed for the problem of multimodal emotion recognition in conversations that are:  (i) performed under unexpected automatic speech recognition errors; (ii) exploring inference in cases of noisy or absent modalities.

\end{document}